\documentclass[onecolumn]{aa}
\usepackage{graphicx}
\begin{document}
   \title{Optical Spectroscopy of BL Lac Objects:\\
new redshifts and mis-identified sources.}

      \author{N. Carangelo, 
          \inst{1,4}
          R. Falomo  \inst{2}, J. Kotilainen  \inst{3}, A. Treves  \inst{4} and M.-H Ulrich \inst{5} }
   \offprints{N. Carangelo}
   \institute{Universit\`a di Milano-Bicocca\\
		Piazza della Scienza 3, 20126 Milano, Italy\\
		\email{nicoletta.carangelo@mib.infn.it}
	\and
		Osservatorio Astronomico di Padova \\
		Vicolo dell'Osservatorio 5, 35122 Padova, Italy\\
		\email{falomo@pd.astro.it}
	\and
		Tuorla Observatory, University of Turku\\ 
		V\"ais\"al\"antie 20,FIN--21500 Piikki\"o, Finland\\
		\email{jarkot@astro.utu.fi}
	\and
		Universit\`a dell'\ Insubria\\
		via Valleggio 11, 22100 Como, Italy \\
		\email{treves@uninsubria.it }
	\and	European Southern Observatory, \\
	        ESO Headquarter Garching,\\
		Karl-Schwarzschild-Str.2\\
		D-85748 Garching bei M\"unchen, Germany\\
	        \email{mhulrich@eso.org}
}

   \date{Received 23 January 2003/ Accepted 12 June 2003}

   \abstract{We are carrying out a multi-purpose program of high signal 
to noise optical
 spectroscopy at  medium resolution of 30 BL Lac objects. Here we report the detection of three new redshifts, and the discovery
of three misclassified sources.
The new redshifts refer to PKS0754+100 (z=0.266), H-1914-194 (z=0.137) and 
 1ES0715-259, for which we derive a  redshift of z=0.465 and also 
a ``new''  
classification a Steep Spectrum Radio Quasar. In two cases 
(UM493 and 1620+103) stellar spectra indicate a wrong classification. 
The three sources
 with new redshift measurement show strong [OIII] emission in their spectra
 and their  host galaxy properties are known. 
 The central black hole masses derived using the M$_{BH}-L_{bulge}$ 
and the M$_{BH}$-$\sigma$([OIII]) relations are compared.

   }
\authorrunning{Carangelo et al.}
	\titlerunning{Optical Spectroscopy of BL Lacs}
   \maketitle
%

\section{Introduction}

BL Lac objects are strong  radio-loud sources that constitute a 
rather rare subclass of active galactic nuclei (AGN), distinguished  by
 peculiar properties. In particular at variance with other classes of
 AGNs, BL Lacs are characterized by a lack of prominent emission lines,
 highly variable non-thermal continuum and strong, variable optical 
polarization. The quasi featureless optical spectra hinder the determination of the 
redshift (and therefore of the distance) for these sources.  To date only
 for half of the possible population of BL Lacs 
($\sim$ 600 objects, see V\'eron C\'etty and V\'eron  2001) is
the redshift known. \\

The redshift can be measured from the weak emission or absorption lines that are expected to come from the underlying galaxy, from the nucleus or, for very distant objects, from intervening material. The detection of these lines requires optical spectroscopy observations of appropriate signal-to-noise ratio and resolution and in particular depends  on the brightness state of the object: during low state the weak non-thermal contribution of the nucleus allows the measurement of stellar lines of the host galaxy, while in the bright state their detection is more difficult, but the higher contrast
renders more feasible the detection of intervening absorptions.\\

We are carrying out a systematic study of the optical spectra of 30 BL Lacs 
observed during two campaigns at the 3.6 m  ESO telescope. The study aims
to  {\bf (a)} measure new redshifts, which is the subject
of the present paper, {\bf (b)} search for broad 
emission lines, {\bf (c)} perform a statistical study of intervening absorption
 systems, and {\bf (d)} measure the host galaxy velocity dispersion $\sigma$ 
 (Treves et al 2003, Falomo et al. 2003b). For point {\bf (a)} we have concentrated on the brightest available targets taken from major multivalength surveys, for which there is no indication of a host galaxy.
Preliminary results have been presented by
Carangelo et al. 2002. The objects of the sample for  which new redshifts or new classifications
were found are reported in Table 1.

The structure of the paper is as follows: in Section 2 we describe the observations and data analysis, in Section 3 we report the results with comments on individual sources. Section 4 gives a summary and discussion of the results.
In our analysis H$_{0}$=50 Km s$^{-1}$ Mpc$^{-1}$ and q$_{0}$=0 were used to facilitate comparison  with literature results (note that all objects have z$<$0.5)\\
 

\section{Observations and Data analysis}

Observations were gathered at La Silla (Chile) using the 3.6m European Southern Observatory telescope equipped with the ESO Faint Object Spectrograph and Camera (EFOSC2) during two  runs in July 2001 (run A) and in January 2002 (run B). We observed in dark conditions and the seeing was 0.6$''$-1.5$''$.
We obtained optical spectra at  $\sim$ 12 {\AA} resolution (FWHM)  in 
the visible (4085-7520 {\AA}) using a grism at 1.68 {\AA}/pixel dispersion 
combined with an 1.2$''$ slit, centered on the object.\\

Standard data reduction was performed using different packages in IRAF\footnote{IRAF is distributed by the National Optical Astronomy Observatories, which are operated by the Association of the Universities for Research in Astronomy, Inc., under cooperative agreement with the National Science Foundation}  in order to obtain 1-dimensional wavelength-calibrated extracted spectra. The spectra were also flux calibrated using several observations of spectrophotometric stars (Oke 1990) obtained on the same night.\\  

For each object we secured 2-3 individual spectra and then combined them in order to improve the signal to noise ratio. This procedure helps to remove bad pixels and spurious features and to check the reliability of detected absorption and/or emission lines. A journal of the observations including the  reached (average) S/N is given in Table 1. \\

\begin{table}
\centering
\begin{tabular}{lllllccl} \hline
Source &	RA 	&    DEC        &       V 	  &   Run  & S/N	& Texp (s)\\ 
(1)	& (2) & (3) & (4) & (5) & (6) & (7)	\\ \hline
1ES 0715-259  & 07 18 04.9& -26 08 11.0 & 18.0  &	B & 70	& 1500 \\
PKS 0754+100  & 07 57 06.6& +09 56 34.9 & 14.5  &	B & 180 & 1200\\
UM 493        & 12 22 06.0& -01 06 38.0   & 16.4   &	B & 60  & 600\\
1620+103      & 16 22 35.1& +10 13 14.0 & 18.0  &     A  & 70   & 1200\\
H-1914-194    & 19 17 45.1& -19 21 38.0 & 18.5  &	A & 180 & 1200\\ \hline
\end{tabular} 
\caption {The Journal of Observations: column (1) gives the name of the BL Lac Object, (2) and (3) the coordinates (2000), (4) the V magnitude of the BL Lac (from the V\`eron-C\`etty and V\`eron catalog)  ,
  (5) the run of observations (RUN A: 24, 25, 26 July 2001; RUN B: 14, 15, 16 January 2002), (6) average signal-to-noise ratio (S/N), (7) the exposure time.}
\end{table}

\section{Results  on individual sources}

{\bf 1ES 0715-259} belongs to the {\em Einstein} Slew Survey (Elvis et al. 1992) and was classified as a BL Lac Object by Perlman et al. (1996).
 They  assumed that a VLA radio source 
 could be identified with the corresponding Slew Survey source if it fell 
within 80$''$ of the X-ray position. However in this case they reported an 
offset in the radio-X-ray position of $>$200 $''$. The VLA radio map of 
1ES 0715-259 (Perlman et al. 1996)
shows a clear FR II morphology.

The MMT spectrum obtained by Perlman et al.(1996) appeared 
featureless which motivated the BL Lac classification on the basis of the 
selection criteria of Stocke et al. (1991). 
The ratio logF$_{x}$/F$_{r}$=-5.2 
places 1ES 0715-259 near the borderline between  radio-selected and X-ray-selected BL Lac objects.
 There are no polarimetric data for this source.  

HST (Urry et al. 2000) clearly resolved 1ES0715-259 into a point source surrounded by a small, rather elongated host galaxy: both disk and de Vaucouleurs models can fit the data giving for the host galaxy an apparent magnitude of R=21.1 and R=20.0 respectively.\\

We took an optical spectrum of the counterpart of the radio source. This
clearly shows strong emission lines (see figure 1) and the identification, as reported in Table 2,  of the [OII], [NeIII], HeII blended with H$_{\gamma}$, H$_{\beta}$, [OIII] N1 and [OIII] N2 yields a redshift $<z>$=0.465$\pm$0.002. On the basis of the line flux F$_{line}$ reported in Table 2, we derive line luminosities of  L$_{H_{\beta}}\sim 4.5 \times 10^{42}$ erg s$^{-1}$ and L$_{[OIII]N2}\sim 3 \times 10^{42}$ erg s$^{-1}$ typical of radio loud quasar, while for BL Lacs typical L$_{H_{\beta}}\sim 10^{40}$-10$^{41}$ erg s$^{-1}$ (e.g. Sitko and Junkkarinen 1985, their Table 2).

From our optical spectroscopy and the radio properties ($\alpha_{r}$=0.73,
 Perlman 2002, private communication) we propose
 that this source is not a BL Lac but a steep spectrum radio quasar (SSRQ). The mis-classification  by Perlman et al. (1996) is likely due to the poor quality of their optical spectrum, although a dramatic variability of the source 
cannot be excluded.\\

{\bf PKS 0754+100} is a highly polarized and variable source that was discovered by Tapia et al. (1977) and classified as a blazar by Angel \& Stockman (1980). A first tentative redshift proposed by Persic and Salucci (1986, z=0.66) was questioned after the detection of the host galaxy by Abraham et al. (1991) and by Falomo (1996), who suggested z$\sim$0.3 on the basis of the host galaxy properties (at z=0.66 the host would be extremely luminous: M$_{R}\simeq$-25).\\

Our spectrum (see figure 1) shows clearly 
 two emission lines at $\lambda$=4717.85 {\AA}  and $\lambda$=6339 {\AA} (see Table 2): identification of these features with [OII](3727 {\AA}) and [OIII] (5007 {\AA}) yields a redshift $<z>$=0.266$\pm$0.001. This is consistent with the preliminary redshift proposed by  Falomo and Ulrich (2000, z$\simeq$0.28).
At z=0.266 the host galaxy luminosity becomes M$_{R}$=-22.9. We note that at the same redshift there is a companion galaxy, 13 $''$ NE, corresponding to a projected distance of $\sim$ 70 kpc (Pesce et al. 1995).\\

{\bf UM 493, 1620+103} These objects are classified to date as BL Lacs in the main catalogs (e.g Hewitt and 
Burbidge (1993), Padovani and Giommi (1995), Veron-Cetty and Veron (2001)). UM 493 was discovered in a thin-prism survey for extragalactic 
 objects of the University of Michigan  (MacAlpine and Williams 1981). As reported in Figure 1 the spectrum of this source shows absorption lines of the Balmer series rest-frame, clearly indicating that UM 493 is a star of spectral type A.\\

 1620+103 is a radio source listed in the Molonglo
 catalogs MC2 and MC3 (Sutton et al. 1974).  The radio position is 
given with a r.m.s. error in R.A. of 0.75 $''$ and of 10$''$ in DEC 
(see Sutton et al. 1974). Hazard \& Murdoch (1977) proposed possible 
identification  with two possible Blue Stellar Objects (BSOs) 
within $\pm$ $\sim$23$''$ of the radio position. Spectrophotometric 
observations for the sources, first by Baldwin et al. (1973), then by
 Smith et al. (1977)  classified the SW object as a star and the NE object 
as a lineless object. Zotov and Tapia (1979) revealed a polarization 
of 6\% for the NE object, pointing to a BL Lac classification. McIlwrath and Stannard (1980) suggested that the NE object
 is an example of a radio-quiet BL Lac. \\

We gathered spectra for both sources (NE and SW) of Hazard and Murdoch (1977) (see figure 1): the NE object clearly shows the typical absorption line rest-frame of a galactic star and the SW object is confirmed to be a galactic star. We definitively conclude  that the NE object is not a BL Lac but a star probably of spectral type F or intermediate A-F. The X-ray source detected by Della Ceca et al. (1990) (A-2 experiment aboard HEAO-1) coincides with the radio source but not with the star. \\

{\bf H-1914-194} is an X-ray-selected BL Lac (XBL) detected during the HEAO-1 Survey (Schwartz et al. 1989; Laurent-Muehleisen et al. 1993). This source was also detected in the radio during the Parkes-MIT-NRAO Survey (Griffith et al. 1994) and the Texas radio Survey (Douglas et al. 1996). Kollgaard et al. (1996) present a deep radio image for H-1914-194, which shows a bright core with a jetlike feature extended to the northwest toward a local brightening in a diffuse halo (see their figure 5a).

The optical counterpart is determined within 3$''$ of the X-ray position: Kollgaard et al. (1996) give the optical position with an error of $\pm$0.3$''$. HST imaging at high resolution (Scarpa et al. 2000, Urry et al. 2000) clearly resolved this source. The average radial profile of the host galaxy is consistent with both a de Vaucouleurs (R$_{host}$=16.9) and a disk model (R$_{host}$=17.9).

In our spectrum we clearly detect both nuclear emission lines and absorption lines from the host galaxy (see figure 1 and Table 2). Identification of these features with [OII], [OIII] N1 and N2, CaII H\&K and G band yields a 
redshift of $<z>$=0.137$\pm$0.001. \\

\begin{figure}
   \centering
   \includegraphics[width=15cm]{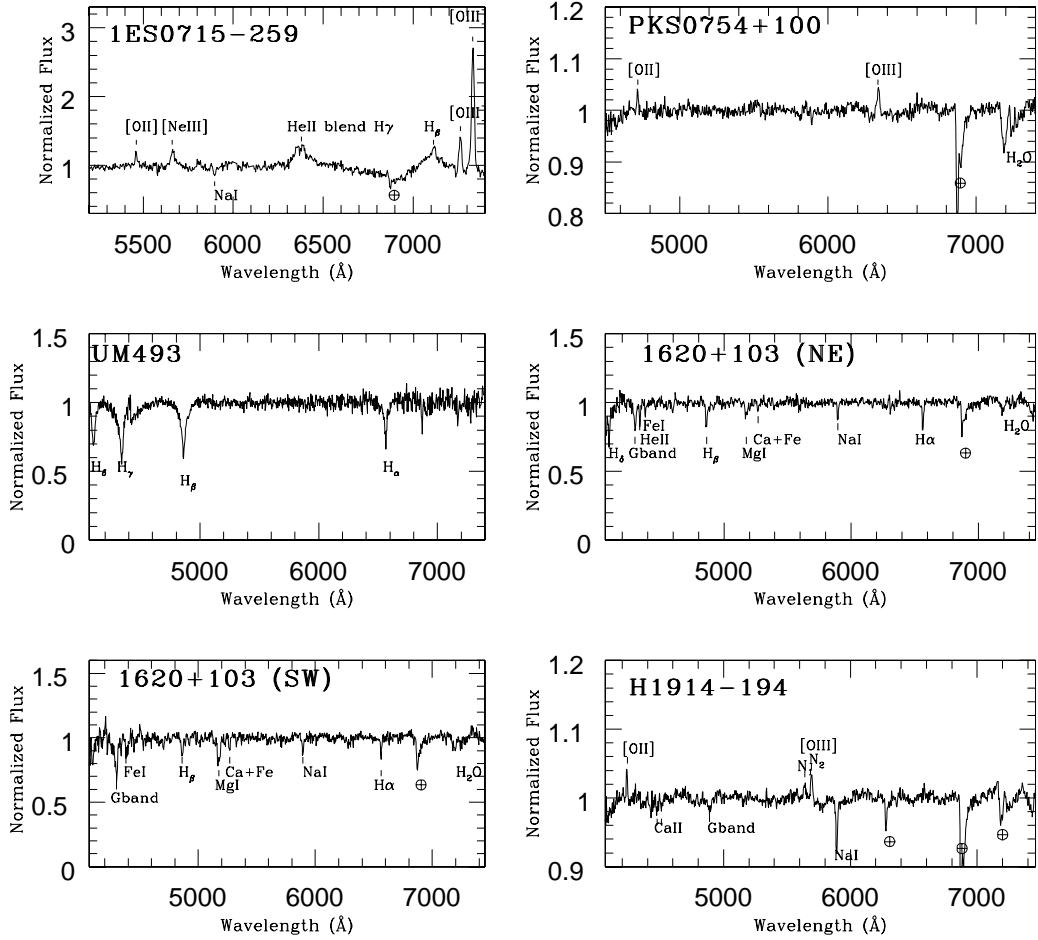}
      \caption{{\bf Normalized optical spectra of the five sources}: for each objects the detected emission and/or absorption lines are marked and identified. We note that in the case of 1ES0715-259 and H1914-194 we also observe an interstellar absorption line at 5892 {\AA} corresponding to NaI.}
    \end{figure}

\begin{table}
\centering
\begin{tabular}{llllllll} \hline
Source 	      &	$\lambda_{obs}$	&	E/A	& Ident.	& z	& EW	& FWHM	&	Flux	\\		
	&	&  &  &    & {\AA} & Km s$^{-1}$ & erg cm$^{-2}$ s$^{-1}$ \\
(1)	& (2)	& (3) & (4)	& (5)	& (6)	& (7)	& (8)	\\ \hline
1ES 0715-259  &   5460.45	& E	& [OII]	& 0.465	& 1.9	& 360	& 1.44$\times$10$^{-16}$	\\
		& 5662.21	& E	& [NeIII] & 0.464 & 4.7 & 820 & 3.29$\times$10$^{-16}$ \\
		& 6353.79	& E	& HeII	 &  0.464 & 2.1 & 1010$^*$ & 1.23$\times$10$^{-16}$ \\
		& 6386.96 	& E 	& H$_{\gamma}$	& 0.471	& 4.1	& 1330$^*$ &2.46$\times$10$^{-16}$ \\
		& 7116.19	& E	&  H$_{\beta}$	& 0.464 & 49.4	& 5200	& 3.20$\times$10$^{-15}$ \\
		& 7262.63	& E 	& [OIII] N1 & 0.465 & 11.9	&	630	& 7.48$\times$10$^{-16}$ \\
		& 7333.04	& E	& [OIII] N2	& 0.465 & 35.4	&  	570	&2.21$\times$10$^{-15}$ \\ \hline
PKS 0754+100  &	4717.85		& E	& [OII]	& 0.266	& 0.6	& 540	& 7.30$\times$10$^{-16}$ \\
		& 6339.00	& E	& [OIII] & 0.266 & 1.1 	& 730	& 1.30$\times$10$^{-15}$ \\ \hline
H-1914-194    &	4240.14	&	E	& [OII]	& 0.138	&	0.5	& 600  	& 3$\times$10$^{-15}$ 	\\
		& 4476.33 &	A	& CaII K & 0.138 & 0.3	& 	& 	\\
		& 4511.04 &	A	& CaII H & 0.137 & 0.2	& 	&		\\
		& 4890.92 &	A	& Gband	& 0.136	& 0.3	& 	&	\\
		& 5639.33 & 	E	& [OIII] N1 & 0.137 & 0.3	& 800 &3.80$\times$10$^{-16}$ \\
		& 5693.55 &	E 	&[OIII] N2	& 0.137	& 0.5	& 690	& 5.40$\times$10$^{-16}$\\ \hline
\end{tabular} 
\caption {Properties of emission and/or absorption lines:
 for each target the observed wavelengths of the detected emission (E) or absorption (A) lines, the identification, the redshift, the equivalent width (EW) and for the emission lines the full width half maximum (FWHM) corrected as described in the text and the flux observed are reported. In two case indicated as $^*$ in the table the FWHM measured are very uncertain because the two emission lines blend.}
\end{table}   

\section{Discussion}

Our results are summarized in Fig.1 and Table 2 where we report for each object the observed wavelengths of the detected emission (E) and/or absorption (A) lines, the identifications, the measured redshift, the equivalent width and for the emission lines the FWHM and the flux observed. In particular we correct the measured FWHM of the detected emission lines taking into account the instrumental resolution of $\sim$420 Km s$^{-1}$, that has been subtracted in quadrature from the measured widths. Note that this correction changes the line width by as much as 15-20 \%.\\
 
 We found for  three sources that the BL Lac classification reported in the literature is wrong. 
In particular, 1ES 0715-259 is clearly a steep spectrum radio-loud quasar at 
a redshift of z=0.465 while UM493 and 1620+103 are two stars.
 For the two confirmed BL Lac objects we measured new redshifts z=0.266 for PKS0754+100 and z=0.137 for H-1914-194.\\

From the measured z 
we can derive the absolute magnitudes of the nuclei and of the host galaxies
based on the  apparent magnitudes measured with HST 
 (Scarpa et al. 2000 and Urry et al. 2000). 
Moreover we note that  the three sources  with determined redshifts show clear [OIII] 
emission in their spectra.\\

It is therefore possible to estimate the mass of the central black hole
in two independent ways. The former is based on the M$_{BH}-L_{bulge}$ relation(e.g. Kormendy \& Gebhardt 2001, Falomo et al. 2003a and references therein). In particular
we have used the procedures described in detail by Bettoni et al (2003).
 The results for the three objects under examination are reported
in Table 3. The uncertainties are $\sim$40\%. An alternative procedure is to start
from the $\sigma$([OIII])-$\sigma^{*}$ (velocity dispersion) relation proposed by Nelson and Whittle (1996), Nelson (2000), Shields et al. (2003), 
Boroson (2003), and to take advantage
of the M$_{BH}$-$\sigma$ relation (Ferrarese \& Merritt 2000; Gebhardt et al. 2000, Merritt \& Ferrarese 2001b). Even if our observations yield only
a rough estimate of the [OIII] width,
 the derived $\sigma$([OIII]) are in the range (170-370  Km s$^{-1}$), 
 typical values of $\sigma^{*}$ measured in BL Lac objects (e.g. Barth et al. 2003, Falomo et al. 2002, 2003b, Treves et al. 2003).   Mainly because of the large
dispersion of the M$_{BH}$-$\sigma$([OIII]) relation, the resulting mass (Table 3)
is more uncertain than with the previous technique, so that the results
are not formally discordant. They indicate the uncertainties (Boroson 2003) and the potentiality  of the method
of determining M$_{BH}$  for BL Lacs at large redshift, 
where the host galaxy is hard to
 detect. \\

\begin{table}
\centering
\begin{tabular}{cccccc} \hline
Source 	       & z 	& M$_{R_{{host}}}$ & log(M$_{BH}/M_{\odot})_{{L_{bulge}}}$	& $\sigma$([OIII]) (km s$^{_1}$) & log(M$_{BH}/M_{\odot})_{\sigma([OIII]}$ \\ 
(1)	& (2) & (3) & (4) & (5) & (6) \\ \hline
1ES 0715-259   & 0.465  & -24.17 & 9.1 & 244 & 8.6 \\
PKS 0754+100   & 0.266  & -22.64 & 8.3 & 312 & 9.1 \\
H-1914-194     & 0.137  & -23.27 & 8.6 & 292 & 8.9 \\ \hline
\end{tabular} 
\caption {The black hole masses derived using the M$_{BH}-L_{bulge}$ and M$_{BH}$-$\sigma$([OIII]) relations: column (1) gives the source name, (2) the redshift measured in this study, (3) the absolute R host galaxy magnitude corrected for extinction, K-correction and evolution, (4) the BH mass derived using the M$_{BH}-L_{bulge}$ relation, (5) the velocity dispersion $\sigma$([OIII]) and (6) the BH mass derived from M$_{BH}$-$\sigma$([OIII] relation.}
\end{table}   
 
\begin{acknowledgements}
We are grateful to Eric Perlman for comments. This research has made use of the NASA/IPAC Extragalactic Database (NED) which is operated by the Jet Propulsion Laboratory, California Institute of Technology, under contract with the National Aeronautics and Space Administration. 
\end{acknowledgements}

\end{document}